\documentclass[reprint, aps, prd, showpacs]{revtex4}

\usepackage[utf8]{inputenc}
\usepackage[T1]{fontenc}
\usepackage{amsmath,amssymb}
\usepackage{hyperref}
\usepackage{graphicx}
\usepackage{subfigure}

\usepackage[usenames]{xcolor}

\begin{document}
\pacs{04.70.Dy, 02.20.Sv, 04.60.Pp}
\title{Quantum black holes in Loop Quantum Gravity} 
\author{Rodolfo Gambini}
\email{rgambini@fisica.edu.uy}
\affiliation{Instituto de F\'{i}sica, Facultad de Ciencias, Igu\'a 4225, esq.\ Mataojo, Montevideo, Uruguay}
\author{Javier Olmedo}
\email{jolmedo@fisica.edu.uy}
\affiliation{Instituto de F\'{i}sica, Facultad de Ciencias, Igu\'a 4225, esq.\ Mataojo, Montevideo, Uruguay}
\author{Jorge Pullin}
\email{pullin@lsu.edu}
\affiliation{Department of Physics and Astronomy, Louisiana State University, Baton Rouge, LA 70803-4001}

\begin{abstract}
  We study the quantization of spherically symmetric vacuum spacetimes
  within loop quantum gravity. In particular, we give additional
  details about our previous work in which we showed that one could
  complete the quantization of the model and that the singularity inside
  black holes is resolved.  Moreover, we consider an alternative
  quantization based on a slightly different kinematical Hilbert
  space. The ambiguity in kinematical spaces stems from how one treats
  the periodicity of one of the classical variables in these
  models. The corresponding physical Hilbert spaces solve the
  diffeomorphism and Hamiltonian constraint but their intrinsic
  structure is radically different depending on the kinematical
  Hilbert space one started from. In both cases there are quantum
  observables that do not have a classical counterpart. However, one
  can show that at the end of the day, by examining Dirac observables,
  both quantizations lead to the same physical predictions.
\end{abstract}

\maketitle

\section{Introduction}

One of the problems that blocks the completion of the quantization
program of loop quantum gravity \cite{LQG} is the suitable
implementation of the dynamics. As is usual in canonical approaches to
gravity, the dynamics is codified in a set of first class constraints
with an algebra involving structure functions, which prevents the
direct application of Dirac quantization approach \cite{dirac}. This
problem is also inherited by some of the symmetry reduced models of
the theory. Although it is absent in the most studied symmetry reduced
models corresponding to homogeneous spacetimes, the so-called loop
quantum cosmology \cite{LQC}, the problem reappears as soon as one
considers models with spatial dependence in the variables. Some of the
most relevant reduced models to be studied, due to their physical
implications, are the spherically symmetric spacetimes. They can for
instance describe the final stage of the gravitational collapse of
astrophysical objects. However, the classical description turns out to
be incomplete since this kind of geometries is characterized by the
presence of a classical singularity.  These reduced models are also
interesting from the technical point of view, since they have a
Hamiltonian constraint representing invariance under time
reparametrizations and a diffeomorphism constraint associated with the
symmetry under redefinitions of the radial coordinate. It is one of
the simplest inhomogeneous reduced model of general relativity.

The first attempts to quantize spherically symmetric vacuum models
\cite{thiem-kast, kuchar} applied standard quantization techniques to the
``old'' (complex) Ashtekar variables and the traditional metric variables
respectively, performed a series of canonical transformations and gauge fixings
and yielded a quantum theory where the physical states of the spacetime
correspond to superpositions of Schwarzschild geometries peaked at a given mass.
There is no sense in which the singularity of the spacetime is eliminated by
the quantization, as it is in loop quantum cosmology. A reanalysis of these
approaches using the ``modern'' (real) Ashtekar variables and performing a loop
quantization yielded essentially the same result \cite{cgp}. It appears that the
unexpected presence of the singularity within these novelty quantization
approaches is due to the ``severe gauge fixing'' adopted for the model before
attempting the quantization and there was too little left in terms of dynamics
for quantization to be able to do anything about the singularity.
  On the other hand, several studies of the interior \cite{interior} of the
Schwarzschild spacetime, exploiting its isometry to the Kantowski--Sachs metric
and treating it with loop quantum cosmology techniques, have indicated that the
singularity is resolved, therefore creating a tension with the previous studies
based on gauge fixing. These are the main motivations to explore alternative
quantization approaches where one quantizes before completely gauge fixing the
model. 

We base our work on previous papers that developed the spherically
symmetric framework. The latter was  already studied adopting 
the complex Ashtekar
variables \cite{bengtsson,thiem-kast} and the Ashtekar-Barbero
connection \cite{bojo-kast}.  A more complete description carried out
in Refs.  \cite{boswi12,boswi} allowed to establish the kinematical
framework together with a prescription for the quantum
constraints. However, the physical states were not provided. In more
recent work \cite{gp-lett} it was shown that a redefinition of the
constraint algebra of the diffeomorphism and Hamiltonian constraint
can be carried out that turns it into a Lie algebra.  One can then
apply the Dirac quantization approach and solve the model
quantum mechanically. The physical space of states was identified and
the metric operator realized as an evolving constant of motion and
shown to be singularity free. Moreover, new quantum Dirac observables
without classical counterpart were identified.

In this paper we give more details about the study of
Ref.~\cite{gp-lett}.  We will also analyze two different quantization
prescriptions stemming from different choices of kinematical Hilbert
space. It turns out that the introduction of a diffeomorphism
invariant inner product in one plus one dimensions admits more than
one kinematical implementation.  The reason for the ambiguity is that
one of the variables of the problem (associated with the direction
transversal to the radial one) is a scalar. As is usual in loop
quantum gravity, when one has scalar variables, one represents them
using point holonomies \cite{Thiemann:1997rq}. That can be done in
two different ways. One could consider the exponentiation of the
variable and therefore the resulting functions are periodic in the
variable. An alternative, which is what is normally done in loop
quantum cosmology, is to consider a Bohr compactification. In
Ref.~\cite{gp-lett} we proceeded in the first way, without introducing
a Bohr compactification. On the other hand, using the Bohr
compactification corresponds better to what is done in the full
theory. One discovers in this case that there exist superselection
sectors associated with the periodicity of the variable. The choice of
different kinematical Hilbert spaces leads to quite different
implementations of the Hamiltonian constraint and to very different
looking spaces of physical states that satisfy all the constraints. So
apparently one is faced with two inequivalent quantizations. However,
a more careful study of the Dirac observables shows that the physical
content of both quantizations is the same.

The paper is organized as follows. In Sec.~\ref{sec:class_sys} we describe
the classical system and we establish the new constraint algebra. The kinematics
quantum description  is provided in Sec.~\ref{sec:kinemat} together with the
physical solutions in Sec.~\ref{sec:scalar-const} and
Sec.~\ref{sec:diffeo-const}. A brief description about semiclassical states is
included in Sec.~\ref{sec:semiclass} and the final conclusions in
\ref{sec:conclus}. Appendix~\ref{app-A} includes the relation with the metric
variables, the falloff conditions and the boundary terms.

\section{Classical system}\label{sec:class_sys}

The reduction of the full theory to spherically symmetric vacuum
geometries was considered in Ref. \cite{Thiemann:1997rq}, based on the
original ideas of Ref. \cite{spheclly}. There, one introduces three
Killing vectors compatible with the spherical symmetry and demands
that the Lie derivatives of the triad and connection along their
orbits is compensated by an internal O(3) transformation. These
conditions, as it was shown explicitly in Ref. \cite{Thiemann:1997rq},
provide the reduced theory for complex Ashtekar variables. Likewise,
the reduction for real Ashtekar variables can be carried out in a
similar fashion (see for instance Refs. \cite{bojo-kast,
  boswi}). Moreover, the same results are obtained if one starts with
the metric variables, reduces the theory to spherical symmetry and
then introduces Ashtekar-like variables for the reduced theory, as was
done for instance in \cite{gps}.

We will then follow the treatment suggested by Bojowald and
Swiderski \cite{boswi} for spherically symmetric spacetimes within
  loop quantum gravity. The Ashtekar variables adapted to a
spherically symmetric spacetime, are given by
\begin{align}\nonumber
 A &=A_{a}^i\tau_i dx^a= A_x(x)\tau_3d x + [A_1(x)\tau_1+A_2(x)\tau_2]d\theta \\
 &+[A_1(x)\tau_2-A_2(x)\tau_1]\sin\theta d\phi+\tau_3\cos\theta d\phi,\\\nonumber
E &=E_{i}^a\tau^i 	\partial_a=\sin\theta \Big( E^x(x)\tau_3 \partial_x + [E^1(x)\tau_1+E^2(x)\tau_2]\partial_\theta\Big)\\
&+[E^1(x)\tau_2-E^2(x)\tau_1]\partial_\phi,
\end{align}
where $\tau_i$ are the generators of $SU(2)$
(i.e. $[\tau_i,\tau_j]={\epsilon_{ij}}^k\tau_k$ with $\epsilon_{ijk}$
the totally antisymmetric tensor), $x$ is a radial coordinate,
  and $\theta\in[0,\pi)$ and $\phi\in[0,2\pi)$ the angular
  coordinates. The reduced Poisson algebra is given by
\begin{align}\nonumber
&\{A_x(x),E^x(x')\}=2 G\gamma \delta(x-x'),\\
&\{A_i(x),E^j(x')\}=G\gamma \delta_i^j\delta(x-x'),\quad i,j=1,2,
\end{align}
where $G$ is the Newton constant and $\gamma$ is the Immirzi
parameter. The spacetime metric is
\begin{equation}\label{eq:1+3metric}
ds^2=-(Ndt)^2+q_{xx}(dx+N_rdt)^2+q_{\theta\theta}d\Omega^2,
\end{equation}
where $N$ and $N_r$ are the lapse and the shift functions, respectively, $t$ is time coordinate and 
$d\Omega^2=d\theta^2+\sin^2\theta d\varphi^2$ is the induced metric in the unit
sphere. In terms of the triad variables, the metric components read $q_{xx}=((E^1)^2+(E^2)^2)/E^x$ and $q_{\theta\theta}=E^x$.

In this situation, we are left with three first class constraints:
  a scalar constraint, a diffeomorphism constraint in the radial
  direction and a remnant Gauss constraint generating $U(1)$ gauge
  transformations (the $SU(2)$ symmetry is broken after the
  reduction).  Moreover, we will adopt the usual description, in
  order to identify the gauge invariant variables of the model,
obtained after several canonical transformations. One first introduces
polar coordinates, i.e.,
\begin{align}\nonumber
E^1&=E^\varphi\cos(\alpha+\beta), &E^2&=E^\varphi\sin(\alpha+\beta),\\
A_1&=A_\varphi\cos\beta, &A_2&=A_\varphi\sin\beta,
\end{align}
and completes the canonical transformation defining 
\begin{align}\nonumber
\eta&=\alpha+\beta, \quad P^\eta=A_\varphi E^\varphi\sin\alpha=2A_1E^2-2A_2E^1,\\
\bar A_\varphi &= 2A_\varphi\cos\alpha.
\end{align}
This transformation together with the redefinition of the $su(2)$ algebra 
\begin{align}\nonumber
\tilde\tau_1(x)&=\tau_1\cos\eta +\tau_2\sin\eta,\\
\tilde\tau_2(x)&=-\tau_1\sin\eta +\tau_2\cos\eta,
\end{align}
diagonalizes the densitized triad in the form
\begin{align}\nonumber
E &= E^x(x)\tau_3\sin\theta \partial_x + E^\varphi(x)\tilde\tau_1(x)\sin\theta \partial_\theta\\
 &+E^\varphi(x)\tilde\tau_2(x)]\partial_\phi. \\\nonumber
\end{align}
The canonical Poisson brackets are now given by
\begin{align}\nonumber
&\{ A_x(x),E^x(x')\}=2 G\gamma \delta(x-x'),\\\nonumber
&\{\bar A_\varphi(x),E^\varphi(x')\}=2G\gamma \delta(x-x'),\\
&\{\eta(x),P^\eta(x')\}=2G\gamma \delta(x-x').
\end{align}
The last canonical transformation
\begin{align}
\bar A_x&=A_x+\eta',\quad  \bar P^\eta = P^\eta+(E^x)',
\end{align}
allows one to simplify the treatment of the pure gauge canonical pair $\eta$ and
$\bar P^\eta$. The connection is finally given by
\begin{align}\label{eq:connection}\nonumber
A &= (\bar A_x-\eta')\tau_3d x +\bar  A_\varphi[\cos\alpha\tilde\tau_1-\sin\alpha\tilde\tau_2]d\theta \\
 &+\bar A_\varphi[\sin\alpha\tilde\tau_1+\cos\alpha\tilde\tau_2]\sin\theta d\phi+\tau_3\cos\theta d\phi.
\end{align}
The spin connection in these variables can be computed straightforwardly, yielding
\begin{align}\nonumber
 \Gamma &=\Gamma_{a}^i\tau_i dx^a= -\eta'\tau_3d x +\frac{(E^x)'}{2E^\varphi} \tilde\tau_2d\theta -\frac{(E^x)'}{2E^\varphi} \tilde\tau_1\sin\theta d\phi+\tau_3\cos\theta d\phi.
\end{align}
Together with the connection $A$ given in Eq.~\eqref{eq:connection}, we can compute the components of the curvature $\gamma K=A-\Gamma$
\begin{align}\label{eq:curvat-to-conn}
2\gamma K_x &=\bar A_x,\quad 2\gamma K_\varphi = \bar A_\varphi,
\end{align}
(in the following, we will assume $\gamma = 1$). Therefore, thanks to
Eq.~\eqref{eq:curvat-to-conn}, one is left at the end of the day with
three pairs of canonical variables $(E^x,K_x)$,
$(E^\varphi,{K}_\varphi)$ and $(\eta,\bar P^\eta)$, whose
  geometrical interpretation is simple: the triad components $E^x$ and
  $E^\varphi$ are related to the metric components ---see
  Eqs. \eqref{eq:metric-to-triad}--- and the variables $K_x$ and
  ${K}_\varphi$ are the components of the extrinsic curvature.

The total Hamiltonian for the theory is a linear combination of the Gauss, diffeomorphism and scalar constraints. 
The contribution of the former to the Hamiltonian  is given by $\int dx\lambda_\phi G_\phi$ where $G_\phi:=\bar P^\eta$ is the Gauss constraint and $\lambda_\phi$ is the corresponding Lagrange multiplier. In the following,
we will consider the gauge fixing $\eta = 0$ in order to eliminate the gauge freedom
corresponding to this constraint. One can in fact  implement consistently
this gauge symmetry at the quantum level \cite{chiou}, however, our purpose is just
simplify the study as much as possible, keeping special attention to other more interesting aspects.
Therefore, we will work with a total reduced Hamiltonian 
\begin{equation}\label{eq:total-ham}
  H_T=G^{-1}\int dx (NH+N_rH_r),
\end{equation}
that is a linear combination of
the diffeomorphism and scalar constraints
\begin{subequations}\label{eq:old-constr}
\begin{align}
& H_r:=E^\varphi K_\varphi'-(E^x)' K_x\,,\label{eq:difeo}\\ \nonumber
  &H :=\frac{\left((E^x)'\right)^2}{8\sqrt{E^x}E^\varphi}
-\frac{E^\varphi}{2\sqrt{E^x}} - 2 K_\varphi \sqrt{E^x} K_x  
-\frac{E^\varphi K_\varphi^2}{2 \sqrt{E^x}}\\
&-\frac{\sqrt{E^x}(E^x)' (E^\varphi)'}{2 (E^\varphi)^2} +
\frac{\sqrt{E^x} (E^x)''}{2 E^\varphi}\,,\label{eq:scalar1}
\end{align}
\end{subequations}
respectively. In addition, we will not include here
contributions on the boundaries for simplicity (see App.~\ref{app-A}).

One can easily check that the constraint algebra is
\begin{subequations}
\begin{align}
&\{H_r(N_r),H_r(\tilde N_r)\}=H_r(N_r\tilde N_r'-N_r'\tilde N_r),\\
&\{H(N),H_r(N_r)\}=H(N_r N'),\\
&\{H(N),H(\tilde N)\}=H_r\left(\frac{E^x}{(E^\varphi)^2}\left[N\tilde N'-N' \tilde N\right]\right),
\end{align}
\end{subequations}
and is equipped with structure functions (like in the general theory), with the
ensuing difficulties for achieving a consistent quantization \cite{haji-kuch}.

Let us emphasize that, in order to recover a Schwarzschild spacetime, one
should  consider suitable fall-off conditions for the fields of the previous
reduced theory. They were already studied in Ref. \cite{kuchar}, and we
summarize  the main aspects in App. \ref{app-A}. In this case, the total action
of the model corresponds to the previous reduced canonical theory plus a global
degree of freedom emerging out of boundary terms, whose contribution can be
identified with the mass of the black hole $M$, and its canonically conjugated
momentum $\tau$, which is the proper time of an observer at infinity.

\subsection{New constraint algebra}

Following previous ideas (see Refs.~\cite{cgp,gp-lett}), we will modify the constraint
algebra in a suitable way by introducing the new lapse $\bar N$ and shift $\bar N_r$ functions, such that
\begin{equation}
N_r=\bar N_r -2 N\frac{K_\varphi\sqrt{E^x}}{\left(E^x\right)'},\quad N = \bar N \frac{\left(E^x\right)'}{E^\varphi}.
\end{equation}
After this transformation, the diffeomorphism constraint~\eqref{eq:difeo} remains unaltered, however, the scalar one~\eqref{eq:scalar1} is now
\begin{equation}
  H =-\int dx \bar N\left[ \sqrt{E^x}\left(1-\frac{[(E^x)']^2 }{4 (E^\varphi)^2}+K_\varphi^2\right)\right]'\,.
\end{equation}
We can integrate the scalar constraint $H(\bar N)$ by parts, which yields
\begin{align}\label{eq:abel-class-contr}
&H(\tilde N) =-\int dx \tilde N
\bigg(-\sqrt{E^x}\left(1+K_\varphi^2\right)+2 G M+\frac{[(E^x)']^2\sqrt{E^x}}{4
    \left(E^\varphi\right)^2}\bigg),
\end{align}
where the new lapse function is now $\tilde N := \bar N'$.
The term $2GM$ emerges after imposing the boundary conditions for
the lapse \cite{kuchar,cgp,gp-lett} (see App.~\ref{app-A}), and it ensures, as we will see below, the existence of Schwarzschild-like solutions.  With this rescaling the Hamiltonian constraint turns out to have an
Abelian algebra with itself, and the usual algebra with the diffeomorphism
constraint
\begin{equation}\label{diff-constr}
H_r(\bar N_r)=\int dx \bar N_r \left[-
(E^x)' K_x +E^\varphi K_\varphi'\right].
\end{equation}
More explicitly,
\begin{subequations}
\begin{align}
&\{H_r(\bar N_r),H_r(\bar M_r)\}=H_r(\bar N_r\bar M_r'-\bar N_r'\bar M_r),\\
&\{H(\tilde N),H_r(\bar  N_r)\}=H(\bar N_r \tilde N'),\\
&\{H(\tilde N),H(\tilde M)\}=0,
\end{align}
\end{subequations}
which turns out to be a Lie algebra, allowing us to apply standard
quantization techniques.

We might notice that the Hamiltonian constraint \eqref{eq:abel-class-contr} can
be written at the classical level as  
\begin{equation}
H(\tilde N)=\int dx \tilde NH_-H_+,
\end{equation}
with
\begin{equation}
H_\pm=\sqrt{\sqrt{E^x}\left(1+K_\varphi^2\right)-2 G M}\pm\frac{\left(E^x\right)'(E^x)^{1/4}}{2
    E^\varphi}.
\end{equation}
Since the interesting physical sectors will be those with $E^x>0$,
$\left(E^x\right)'>0$ and $E^\varphi>0$, the vanishing of $H(N)$ corresponds in fact to
$H_-=0$. Therefore, after the redefinition of the lapse function $\underline{N}=\tilde N H_+/(2E^\varphi)$, we get
the constraint
\begin{align}\label{eq:scalar-cons-A}
&H(\underline{N})=\int dx \underline{N}\bigg(2E^\varphi\sqrt{\sqrt{E^x}\left(1+K_\varphi^2\right)-2 G M}-\left(E^x\right)'(E^x)^{1/4}\bigg).
\end{align}
that is classically equivalent to \eqref{eq:abel-class-contr}. This last form of
the Hamiltonian constraint will be more suitable for dealing with the
quantization in which only periodic functions of the point holonomies are
considered.

Let us remark that the sets of constraints \eqref{eq:abel-class-contr} and
\eqref{diff-constr}, and \eqref{eq:old-constr} lead to the same metric solution:
the Schwarzschild metric. This is a direct consequence of the fact that the new
scalar constraint is actually a linear combination (with coefficients dependent
on the dynamical variables) of the old scalar and diffeomorphism constraints.
Concretely, for the exterior, the gauge conditions $E^x=x^2$ and $K_\varphi=0$,
their conservation upon evolution and the restriction to the constraint surface
yields $N_r=0$ and
\begin{equation}
E^\varphi=\frac{x}{\sqrt{1-\frac{2GM}{x}}},\quad N=1-\frac{2GM}{x}.
\end{equation}
One can straightforwardly realize that the metric components $q_{xx}=(E^\varphi)^2/E^x$ and $q_{\theta\theta}=E^x$ in Eq.~\eqref{eq:1+3metric}, together with these results, allow one to recover the Schwarzschild metric.

\section{Quantization: kinematical structure}\label{sec:kinemat}

In order to start with the quantization, we will adopt a standard
description ${\cal H}_{\rm kin}^m=L^2(\mathbb{R},dM)$ for the global degree of
freedom corresponding to the mass of the black hole. For the remaining ones, we
will consider two kinematical Hilbert spaces whose structure is inherited from  loop quantum gravity
\cite{boswi,chiou}: periodic and quasiperiodic
functions of the point holonomies. In particular, one starts with the space of linear
combinations of holonomies of $su(2)$-connections along edges $e$, or in other
words, cylindrical functions of the  connections through holonomies along the
mentioned edges. We can then introduce the notion of graph $g$ for this reduced
model, which  consists of a collection of edges $e_j$ connecting the vertices
$v_j$. It is natural to associate the variable $K_x$ with non-overlapping edges
along the radial direction in the graph and the scalar $K_\varphi$ with vertices
on it (point holonomies). We will also consider that each edge is connected with
another by means of the corresponding vertex. In general, a given graph can be written as a linear combination of products of cylindrical functions of the form
\begin{align}
&T_{g,\vec{k},\vec{\mu}}(K_x,K_\varphi) =\prod_{e_j\in g}
\exp\left(\frac{i}{2} k_{j} \int_{e_j} dx\,K_x(x)\right)
\prod_{v_j\in g}
\exp\left(\frac{i}{2}  \mu_{j} K_\varphi(v_j) \right),
\end{align}
where the label  $k_j\in\mathbb{Z}$ is the valence associated with the edge $e_j$, and
$\mu_j\in\mathbb{R}$ the valence associated with the vertex $v_j$ (usually
called ``coloring''). 

\subsection{Prescription A: periodic functions of point holonomies}

If one adopts a representation for the point holonomies 
as periodic functions, the labels $\mu_j$ are real but must belong to a countable subset of the real line
with equally displaced points. In this case, the labels $\vec{\mu}$ can be relabeled by an integer, for instance, $\vec{n}$. The kinematical Hilbert space turns out to be
\begin{equation}
{\cal H}^A_{\rm kin}={\cal H}_{\rm kin}^m\otimes\left[\bigotimes_{j=1}^V\ell^2_j\otimes\ell^2_j\right],
\end{equation} 
where $\ell^2$ denotes the space of square
summable functions. It is equipped with the inner product
\begin{align}\label{eq:kin-inner-prod-p}
\langle g,\vec{k},\vec{n},M
|g',\vec{k}',\vec{n}',M'
\rangle=\delta(M-M')\delta_{\vec{k},\vec{k}'}\delta_{\vec{n},\vec{n}'}\delta_{g,g'}\;,
\end{align}
where $\delta_{g,g'}$ is equal to the unit if $g = g'$ or zero otherwise, and with $n_j\in\mathbb{Z}$. This kinematical
Hilbert space is then separable.

\subsection{Prescription B: quasiperiodic functions}

An alternative choice for the kinematical structure closer to the full theory is
for point holonomies represented as quasiperiodic functions of the connection,
as is usually done in loop quantum cosmology, whose kinematical Hilbert space is
$L^2(\mathbb{R}_{\rm Bohr},d\mu_{\rm Bohr})$, with
$\mathbb{R}_{\rm Bohr}$ the Bohr compactification of the real line and $d\mu_{\rm
Bohr}$ the natural translationally invariant measure on that set. In this situation, the kinematical Hilbert
space ${\cal H}^B_{\rm kin}$ turns out to be the tensor product
\begin{equation}
{\cal H}^B_{\rm kin}={\cal H}_{\rm kin}^m\otimes\left[\bigotimes_{j=1}^V\ell^2_j\otimes L_j^2(\mathbb{R}_{\rm Bohr},d\mu_{\rm Bohr})\right],
\end{equation}
which is endowed with the inner product
\begin{align}\label{eq:kin-inner-prod-qp}
\langle g,\vec{k},\vec{\mu},M
|g',\vec{k}',\vec{\mu}',M'
\rangle=\delta(M-M')\delta_{\vec{k},\vec{k}'}\delta_{\vec{\mu},\vec{\mu}'}\delta_{g,g'}\;.
\end{align}
Since the labels $\vec{\mu}$ can take any real value, this kinematical Hilbert
space ${\cal H}^B_{\rm kin}$ is nonseparable, which is the main difference with respect to the previous Hilbert space ${\cal H}^A_{\rm kin}$.

\subsection{Operator representation}

The representation of the basic operators is essentially the same in both
quantizations. In particular, the mass and the triads act as multiplicative
operators on these states
\begin{align}
&{\hat{M} } |g,\vec{k},\vec{\mu},M\rangle
= M |g,\vec{k},\vec{\mu},M\rangle,\\
&{\hat{E}^x(x) } |g,\vec{k},\vec{\mu},M\rangle
= \ell_{\rm Pl}^2 k_j |g,\vec{k},\vec{\mu},M\rangle,
\\
&\hat{E}^\varphi(x) |g,\vec{k},\vec{\mu},M\rangle
= \ell_{\rm Pl}^2 \sum_{v_j\in g} \delta\big(x-x_j\big)\mu_j 
|g,\vec{k},\vec{\mu},M\rangle,
\end{align}
where $k_j$ is either the color of the edge including the point $x\in e_j$ or, if
$x$ is at a vertex,  the color of the edge to the right of the vertex. Besides,
$x_j$ is the position of the vertex $v_j$, with $j=1,2,\ldots$ We also must understand 
$\vec{\mu}$ as a general label which will take the corresponding values in either ${\cal H}^A_{\rm kin}$ or ${\cal H}^B_{\rm kin}$.

Regarding the only connection component $A^\varphi(x)$ that is present in the
scalar constraint, the representation adopted will be in terms of point
holonomies of length $\rho$. The basic operators associated with are
\begin{equation}
N^\varphi_{\pm n\rho}(x) |g,\vec{k},\vec{\mu},M\rangle
= |g,\vec{k},\vec{\mu}'_{\pm n\rho},M\rangle ,\quad n\in \mathbb{N},
\end{equation}
where the new vector $\vec{\mu}'_{\pm n\rho}$ either has just the same components than
$\vec{\mu}$ up to  $\mu_j\to\mu_j\pm n\rho$ if $x$ coincides with a vertex of the graph located at $x_j$, or 
$\vec{\mu}'_{\pm n\rho}$ will be $\vec{\mu}$ with a new component 
$\{\ldots,\mu_j,\pm n\rho,\mu_{j+1},\ldots\}$ with $x_{j}<x<x_{j+1}$. 

We can also construct at the kinematical level the volume operator, given by
\begin{equation}\label{eq:phys-vol}
\hat {\cal V}|g,\vec{k},\vec{\mu},M\rangle =4\pi \ell_{\rm Pl}^3\sum_{v_j\in g} \mu_j \sqrt{k_j} |g,\vec{k},\vec{\mu},M\rangle.
\end{equation}

\section{Representation of the Hamiltonian Constraint}\label{sec:scalar-const}

The complete physical Hilbert space can be determined after identifying the
solutions  to both the scalar and diffeomorphism constraints. In this section we
will study the solutions of the scalar constraint. We will represent it as a
quantum operator and we will find its solutions together with a suitable inner
product for them. We will follow two different quantization prescriptions based
on the previous kinematical Hilbert spaces.

\subsection{Prescription A}\label{sec:prescA}

\subsubsection{The physical Hilbert space}

We will start representing the Hamiltonian constraint given by
\eqref{eq:scalar-cons-A} in the kinematical Hilbert space and determining its
solutions, from which the physical Hilbert space will be constructed. We could
have considered \eqref{eq:abel-class-contr} instead of \eqref{eq:scalar-cons-A}.
At the end of the day they would have given the same physical results, but the
latter is more easily solvable.

In particular, we polymerize the connection components and then the scalar constraint is promoted to a
quantum operator with a suitable factor ordering. Let us restrict the study to
a particular state $\Psi_g$ and positive masses, i.e.
\begin{align}\label{gen-kin-state}
&(\Psi_g|=\int_{0}^\infty dM\prod_{v_j\in g}\int_0^{\pi/\rho}dK_\varphi(v_j)\sum_{\vec k}\langle \vec{k},\vec{K}_\varphi,M|\psi(M)\chi(\vec k)\phi(\vec k;\vec{K}_\varphi;M),
\end{align}
and impose that it be a solution to the scalar constraint 
\begin{equation}
(\Psi_g|\hat H(N)^\dagger=(\Psi_g|\sum_{v_j\in g} N(v_j)\hat H_j^\dagger=0.
\end{equation}
This last condition is equivalent ---up to a global factor $(-1)(\ell_{\rm Pl}^2k_j)^{1/4}$--- to a set of partial differential equations for each vertex $v_j$, with $j=1,\cdots,V$, of the form 
\begin{align}\label{eq:diff-eq-prescA}
4i\ell_{\rm Pl}^2\frac{\sqrt{1+m_j^2\sin^2 y_j}}{m_j}\partial_{y_j}\phi_j+\ell_{\rm Pl}^2(k_j-k_{j-1})\phi_j=0,
\end{align}
with $y_j:=\rho K_\varphi(v_j)$, $\phi$ has been decomposed as the product over the vertices of factors of the form $\phi_j=\phi_j(k_j,k_{j-1},y_j,M)$ and
\begin{equation}
m_j^2=\rho^{-2}\left(1-\frac{2 G M}{\sqrt{\ell_{\rm Pl}^2k_j}}\right)^{-1}.
\end{equation}

The solutions to this set of differential equations are of the form
\begin{equation}
\phi_j=\exp\left\{\frac{i}{4}m_j(k_j-k_{j-1})F(y_j,im_j)\right\},
\end{equation}
with
\begin{equation}
F(\phi,k)=\int_0^\phi\frac{1}{\sqrt{1-k^2\sin^2t}}dt,
\end{equation}
the Jacobi elliptic integral of the first kind. We then conclude that the solutions to the scalar constraint must be
\begin{align}\label{eq:sol-prescA}
&\Psi(\vec{k};\vec{K}_\varphi;M)=\psi(M)\prod_{v_j\in g}\chi(k_j) \phi_j\Big(k_j,k_{j-1},K_\varphi(v_j),M\Big)
\end{align}
where
\begin{align}
&\phi(k_j,k_{j-1},K_\varphi(v_j),M)=\exp\left\{\frac{i}{4}m_j(k_j-k_{j-1})F\Big(\rho K_\varphi(v_j),im_j\Big)\right\},
\end{align}
$\psi(M)$ is the analog wave function of Kucha\v{r}'s proposal~\cite{kuchar} and $\chi(k_j)$ are arbitrary
functions of finite norm on the kinematical Hilbert space.

We may notice that the sign of $m_j$ can change depending on the values of the quantum numbers $M$ and $k_j$. In the following, we will identify the exterior region with real $m_j$ and the interior with pure imaginary $m_j$. Now,
for the exterior of the black hole, i.e, whenever $m_j$ are
real, the functions $\phi_j$ are pure phases. Therefore the states
\eqref{eq:sol-prescA} belong to the kinematical Hilbert space ${\cal H}^A_{\rm kin}$ instead of being distributions, which is the usual situation.
However, in the interior $m_j$ becomes a pure imaginary number. Let us consider
$m_j^2<0$ and finite (in particular $\rho$ must be a non-vanishing, positive
real number). The Jacobi elliptic integral is now
\begin{equation}\label{eq:jacobi-int}
F\Big(y_j,|m_j|\Big)=\int_0^{y_j}\frac{1}{\sqrt{1-|m_j^2|\sin^2t}}dt.
\end{equation}
with the argument $y_j\in(0,\pi]$ (or equivalently $K_\varphi(v_j)\in
(0,\pi/\rho]$). If $|m_j|<1$, this integral is real, and therefore we are in a
similar situation as before: the functions $\phi_j$ are just phases.
Otherwise, i.e. $|m_j|>1$, we have to analyze the problem carefully. 
 In this case one can split the Jacobi elliptic integral \eqref{eq:jacobi-int}
in the sum of three contributions:
\begin{itemize}
\item[i)] For $y_j\in(0,\arcsin m_j^{-1})$, the corresponding contribution will
be called $F_1\Big(y_j,|m_j|\Big)$ and remains real and finite. Then the
functions  $\phi_j$ associated to them are pure phases.
\item[ii)] When $y_j\in(\arcsin m_j^{-1},\pi-\arcsin m_j^{-1})$ the Jacobi
elliptic integral is 
\begin{align}\nonumber
&F\big(y_j,|m_j|\big)=F_1\big(\arcsin m_j^{-1},|m_j|\big)+F_2\big(y_j,|m_j|\big).
\end{align}
 
It has a constant real contribution $F_1$ and an imaginary counterpart $F_2\Big(y_j,|m_j|\Big)$
since in the integrand the square root becomes negative,
so that the corresponding $\phi_j$ are not pure phases. However, they are
bounded, and therefore the functions $\phi_j$ are finite. 

\item[iii)]
Finally, if $y_j\in(\pi-\arcsin m_j^{-1},\pi]$ the Jacobi elliptic integral is now
\begin{align}\nonumber
&F\big(y_j,|m_j|\big)=F_1\big(\arcsin m_j^{-1},im_j\big)\\\nonumber
&+F_2\big(\pi-\arcsin m_j^{-1},|m_j|\big)+F_3\big(y_j,|m_j|\big).
\end{align}
In this interval the argument of the square root in $F_3\big(y_j,|m_j|\big).$ is
positive, and $\phi_j$ becomes a phase that varies with $y_j$.
\end{itemize}
In summary, we conclude that the solutions to the constraint of the
form \eqref{eq:sol-prescA} are well defined for both the exterior and
the interior of the black hole. In particular, they belong to the
kinematical Hilbert space ${\cal H}^A_{\rm kin}$ in the sense that
they have finite norm with respect to the inner product
\eqref{eq:kin-inner-prod-p}. We may notice that we have not required
any self-adjointness condition to the constraint (see App.~\ref{app-B}
for additional details). Nevertheless, it is not a serious obstacle in
the sense that we can still find the physical states which codifies
the dynamics of the quantum system, in agreement with
Ref.~\cite{haji-kuch}. These solutions (and the ones provided in
  Sec.~\ref{sec:prescB}) generically are not associated with a
  semiclassical geometry. However, either making special choices of
  the values of the labels of the states, or considering
  superpositions of states, one can approximate semiclassical
  geometries well. We will discuss this in section
  \ref{sec:semiclass}.

Furthermore, as we will see in Sec.~\ref{sec:diffeo-const}, if one considers all
possible superpositions of these solutions $\Psi_g$ within a diffeomorphism
class of graphs $[g]$, the resulting state $\Psi_{\rm phys}$ will be an element
of ${\cal H}_{\rm Diff}$. We then conclude that the physical Hilbert space is a
subspace of ${\cal H}_{\rm Diff}$ (see Sec.~\ref{sec:diffeo-const} for
additional details) whose inner product is given in Eq.~\eqref{eq:ga-inner-prod}. Therefore, this quantization prescription does not
require the introduction of additional structures (like a different inner
product) with respect  to ${\cal H}_{\rm Diff}$ in order to reach a consistent
quantum description.

\subsubsection{Observables}

In agreement with previous quantizations \cite{kuchar,cgp,gp-lett}, there exists
a Dirac observable corresponding to the mass of the black hole, i.e. $\hat M$,
which can be identified as an observable on the boundary. Therefore, any
physical state will be a linear superposition of black holes with well defined
masses. However, the quantization we present here provides a genuine observable
owing to quantum geometry effects, as was noticed in Ref.~\cite{gp-lett}. Let us
recall that one can select a basis of physical states with a fixed number $V$
of vertices located at $x_j$, with $j=0,1,\ldots,V$. This number $V$ is
preserved under the action of the constraints, then we can construct the
corresponding Dirac observable $\hat V$ that acting on the states $\Psi_g$ gives
the integer number $V$. This observable has no classical analog, and it can be
considered as an observable in the bulk.
 
Moreover, the restriction to invertible diffeomorphisms in the classical theory
leads to identify the triad component $E^x$ with monotonically growing
functions. This condition suggests the consideration of states with
\begin{equation}
\chi(k_{j'})=\delta_{j'j},
\end{equation}
at each vertex $v_j$, respectively, and with monotonically growing integers
$k_j$ at the quantum level, which at the end of the day characterize the
sequence of radii of the geometry. If we also take into account that the  order
of the vertices is also preserved for diffeomorphism invariant states, we can
identify another observable $\hat O(z)$ with $z\in [0, 1]$ in the bulk, such
that
\begin{equation}\label{eq:new-obser}
\hat O(z)\Psi_{\rm phys}=\ell_{\rm Pl}^2 k_{{\rm Int}(Vz)}\Psi_{\rm phys},
\end{equation}
where ${\rm Int}(Vz)$ is the integer part of $V z$.  

Surprisingly, this observable allows us to construct an evolving constant
associated to $E^x$. Given an arbitrary monotonic function $z(x): [0,
x]\to[0, 1]$, the mentioned evolving constant can be constructed as
\begin{equation}
\hat E^x(x)\Psi_{\rm phys}=\hat O\big(z(x)\big)\Psi_{\rm phys}.
\end{equation}
It is clear that freedom in the choice of the function $z(x)$ codifies the gauge
freedom in $E^x$.

\subsubsection{Singularity resolution}\label{sec:sing-resl-presc-A}

One of the questions that one can ask is whether the classical singularity can
be avoided within  this model. The answer is in the affirmative, and in fact,
one can follow two strategies. The first one concerns the requirement of
selfadjointness to the metric components. For instance, the classical quantity
\begin{equation}
g_{tx}=-\frac{(E^x)'K_\varphi}{
    2\sqrt{E^x}}\frac{1}{\sqrt{1+K_\varphi^2-\frac{2 G M}{\sqrt{E^x}}}},
\end{equation}
defined as an evolving constant (i.e. a Dirac observable), must
correspond to a selfadjoint operator at the quantum
level. Classically, $K_\varphi$ and $E^x$ are pure gauge, and $g_{tx}$
is just a function of the observable $M$. The exterior of the black
hole can be covered by the conditions $K_\varphi=0$ and $z(x)=x^2$
(together with an appropriate choice of the lapse and the shift),
leading to the usual form of the Schwarzchild
metric. Quantum-mechanically, the gauge freedom is encoded in
$z(x)$ and the periodic gauge parameter ${\cal K}_\varphi$ (we introduce here
  this calligraphic symbol in order to distinguish between the quantum
  gauge parameter and the classical one $K_\varphi$); its periodicity prevents coordinate singularities. The
physical information is codified by $\hat M$ and $\hat
  O\big(z\big)$. In the interior of the horizon, if $\hat g_{tx}$ is
a selfadjoint operator, a necessary condition will be
\begin{equation}
{1+{\cal K}_\varphi^2 -\frac{2 G M}{\sqrt{\ell_{\rm Pl}^2 k_{j}}}\geq 0.}
\end{equation}
At the singularity, i.e. $j=1$, and owing to the bounded nature of ${\cal K}_\varphi^2<\infty$, 
\begin{equation}
\sqrt{k_1}\geq \frac{2 G M}{\ell_{\rm Pl}(1+{\cal K}_\varphi^2)}>0.
\end{equation}

Therefore, this argument strongly suggests that the classical singularity will
be resolved at the quantum level since $k_1$ must be a non-vanishing
  integer. This truncation of the Hilbert space is consistent because
the action of the constraints does not lead outside the space of
non-vanishing $k$'s. One can therefore analytically continue the
solution to negative values of $x$ and one will have a region of
spacetime isometric to the exterior of the black hole beyond where
the classical singularity used to be.

We will now proceed to present a different quantization prescription, where in
particular the classical singularity can be resolved following alternative
reasonings already employed in the literature~\cite{biancI,mmo}.

\subsection{Prescription B}\label{sec:prescB}

\subsubsection{Hamiltonian constraint}

We will now deal with the solutions to the scalar constraint for the alternate
kinematical Hilbert space of quasi-periodic functions of the point holonomies.
For it we will adopt an alternate prescription for promoting the Hamiltonian
constraint to a quantum operator. Following the usual strategy in loop quantum
cosmology, the scalar constraint corresponds essentially to a difference
operator that only mixes states with support in lattices of constant step (i.e.,
separable subspaces of the kinematical one). However, in order to simplify the
analysis, it is more convenient to adopt a prescription as simple as possible
while it captures all the relevant physical information. On the one hand, we start
by polymerizing the connection $K_\varphi\to\sin\left(\rho
K_\varphi\right)/\rho$ contained in Eq.~\eqref{eq:abel-class-contr}. On the
other hand, the factor ordering ambiguity introduces a freedom in the choice
quantum scalar constraint, that we will take advantage of, picking out a factor
ordering such that: i) the scalar constraint allows us to decouple the zero
volume states, and ii) the different orientations of the triad $E^\varphi$ are
decoupled. It is well known that such features turn out to simplify the
subsequent treatment of the scalar constraint \cite{mmo}. Finally, since there are
inverse triad contributions in Eq.~\eqref{eq:abel-class-contr}, we will adopt
the standard treatment for them by means of the so-called Thiemann's trick
\cite{inv-thiem}, which basically consists in defining them at the classical
level by means of Poisson brackets of certain power of the triad with its
canonically conjugated momentum, and then promote them to quantum commutators
(with the addition that one of the variables may be conveniently polymerized). A
factor ordering that fulfills all the previous requirements is
\begin{align}\label{eq:quant-scalar-constr}
&  \hat{H}(N)=\int dx N(x)\sqrt{\hat{E}^x}\left(  
\hat\Theta\sqrt{\hat{E}^x}+\hat{E}^\varphi\sqrt{\hat{E}^x}
-\frac{1}{4}\widehat{\left[\frac1{\hat{E}^\varphi}\right]} \left[(\hat{E}^x)'\right]^2\sqrt{\hat{E}^x}
-2 G \hat M\hat{E}^\varphi \right),
\end{align}
where the operator $\hat\Theta(x)$ acting on the kinematical states 
\begin{align}
&\hat\Theta(x)|g,\vec{k},\vec{\mu},M\rangle
= \sum_{v_j\in g} \delta(x-x(v_j)) \hat\Omega_\varphi^2 (v_j)
|g,\vec{k},\vec{\mu},M\rangle,
\end{align}
is defined by means of the non-diagonal operator
\begin{align}
&\hat{\Omega}_\varphi (v_j)= \frac{1}{{4i\rho}}|\hat{E}^\varphi|^{1/4}\big[\widehat{{\rm sgn}(E^\varphi)}\big(\hat N^\varphi_{2\rho}-\hat N^\varphi_{-2\rho}\big)+\big(\hat N^\varphi_{2\rho}-\hat N^\varphi_{-2\rho}\big)\widehat{{\rm sgn}(E^\varphi)}\big]|\hat{E}^\varphi|^{1/4}\Big|_{v_j},
\end{align}
where
\begin{equation}
|\hat{E}^\varphi|^{1/4}(v_j)  |g,\vec{k},\vec{\mu},M\rangle
= \ell_{\rm Pl}^{1/2}  |\mu_j|^{1/4}
 |g,\vec{k},\vec{\mu},M\rangle,
\end{equation}
\begin{equation}
\widehat{{\rm sgn}\big(E^\varphi(v_j)\big)}  |g,\vec{k},\vec{\mu},M\rangle
= {\rm sgn}(\mu_j)
 |g,\vec{k},\vec{\mu},M\rangle,
\end{equation}
have been constructed by means of the spectral decomposition of $\hat{E}^\varphi$ on
${\cal H}_{\rm kin}^B$. This choice of the scalar constraint, concretely the operator
$\hat{\Omega}_\varphi$, is well motivated by previous studies of different
cosmological scenarios \cite{biancI,mmo}, owing to the particular structure of
the subspaces invariant under its action, which will be classified below. Regarding the operator 
$\widehat{\left[1/\hat{E}^\varphi\right]}$, 
it will be defined following the mentioned Thiemann's ideas \cite{inv-thiem}, yielding well
defined operators on the kinematical Hilbert space. More precisely, the classical identity
\begin{align}
\frac{{\rm sgn}(E^\varphi)}{\sqrt{|E^\varphi}|}=\frac{2}{G}\{K_\varphi,\sqrt{E^\varphi}\},
\end{align}
with $\{\cdot,\cdot\}$ the classical Poisson brackets, is promoted to a quantum operator (recalling that the connection must be conveniently polymerized). Its square allows us to define
\begin{align}
&\widehat{\left[\frac1{\hat{E}^\varphi}\right]} |g,\vec{k},\vec{\mu},M\rangle
=\sum_{v_j\in g} \delta(x-x(v_j))\frac{{\rm sgn}(\mu_j)}{\ell_{\rm Pl}^2\rho^2}(|\mu_j+\rho|^{1/2}-|\mu_j-\rho|^{1/2})^2
 |g,\vec{k},\vec{\mu},M\rangle.
\end{align}
With all the previous definitions, the action of the Hamiltonian constraint can be computed, yielding
\begin{align}\nonumber
&  \hat{H}(N) |g,\vec{k},\vec{\mu},M\rangle=\sum_{v_j\in g} N(x_j)  (\ell_{\rm Pl}^3k_j)
\left[f_0(\mu_j,k_j,M) |g,\vec{k},\vec{\mu},M\rangle\right.\\
&\left.-f_+(\mu_j)|g,\vec{k},\vec{\mu}_{+4\rho_j},M\rangle-f_-(\mu_j)|g,\vec{k},\vec{\mu}_{-4\rho_j},M\rangle\right],
\end{align}
with the functions 
\begin{widetext}
\begin{align}
&f_\pm(\mu_j)=\frac{1}{16\rho^2}|\mu_j|^{1/4}|\mu_j\pm 2\rho|^{1/2}|\mu_j\pm 4\rho|^{1/4}s_{\pm}(\mu_j)s_{\pm}(\mu_j\pm 2\rho),\\\nonumber
&f_0(\mu_j,k_j,k_{j-1},M)=\mu_j\left(1-\frac{2GM}{\ell_{\rm Pl}|k_j|^{1/2}}\right)+\frac{1}{16\rho^2}\left[{(|\mu_j||\mu_j+ 2\rho|)^{1/2}}s_+(\mu_j)s_-(\mu_j+2\rho)\right.\\
&\left.+{|\mu_j||\mu_j- 2\rho|)^{1/2}}s_-(\mu_j)s_+(\mu_j-2\rho)\right]-\frac{{\rm sgn}(\mu_j)}{\rho^2}(k_j-k_{j-1})^2(|\mu_j+\rho|^{1/2}-|\mu_j-\rho|^{1/2})^2,
\end{align}
\end{widetext}
and where the factors
\begin{equation}
s_{\pm}(\mu_j)={\rm sgn}(\mu_j)+{\rm sgn}(\mu_j\pm 2\rho),
\end{equation}
come from the sign functions incorporated in the definition of the quantum operator $\hat{\Omega}_\varphi$.

\subsubsection{Singularity resolution}

The operator corresponding to the Hamiltonian constraint has been chosen such
that it allows for a singularity resolution. Let us consider the set of
spin networks with $\mu_j>0$ and $k_j>0$, and with an arbitrary number of
vertices. One can easily check that this subspace is preserved under the action
of the scalar constraint (it is, in consequence, an invariant domain).
Concretely, at a given vertex, it preserves $k_j$ and mixes different values of
$\mu_j$ without approaching $\mu_j=0$. The immediate consequence is that one can
construct nontrivial solutions to the Hamiltonian constraint such that they do not have
contributions on states with either $k_j=0$ and/or $\mu_j=0$.
Therefore, the triad components cannot vanish on the space of solutions,
and the classical singularity will not be present in the quantum theory. 
However, one could even think about extending
the previous invariant domain to a bigger one just by adding spin networks with,
e.g., $k_j=0$. In this situation, one can still invoke the arguments shown in
Sec.~\ref{sec:sing-resl-presc-A} about the selfadjointness of some quantum evolving
constants, which will allow one to recover the original invariant domain. Therefore,
one can combine these two procedures in order to achieve a resolution of the
singularity.

In general, one would expect that for the physical states there will be no well
defined notion of horizon or black hole interior, since there is no 
semiclassical geometry associated with them. In those cases it will be difficult to
provide a clear notion of singularity either.  However, one of the most
interesting  situations are  the "most disfavorable" cases (from the point of
view of the existence of singularities) in which one indeed can have a good
approximation of a classical geometry of a black hole throughout most of the
spacetime, and therefore one can check  what happens in the region close to
where the classical singularity would have been. In that case we showed above that
the singularity is eliminated, and the effective geometry becomes regular (and 
discrete) throughout the spacetime.

\subsubsection{Discrete geometry}

The action of the constraint on this orthogonal complement does not mix
different graphs $g$. In other words, the subspace associated with a given graph
$g$ is preserved by the action of the scalar constraint in the sense that no new
vertices are created. In turn, a given graph is partially characterized by the
number of vertices and the set $\{k_j\}$, which is preserved by $\hat{H}(N)$.
Regarding the color of the vertices, the action of the constraint mixes them by
means of a difference operator of step $4\rho$ in the labels~$\mu_j$.

We will assume that any state annihilated by the constraint will belong to the
algebraic dual of the dense subspace $\rm Cyl$ on the kinematical Hilbert space. For a generic graph $g$, it will be of the form
\begin{align}\label{gen-kin-stateB}
&(\Psi_g|=\int_0^\infty dM\sum_{\vec k}\sum_{\vec \mu}\langle g, \vec{k},\vec{\mu},M|\psi(M)\chi(\vec k)\phi(\vec k;\vec{\mu};M). 
\end{align}
It satisfies the constraint equation
\begin{equation}
\sum_{v_j\in g}(\Psi_g|\hat{H}(N_j)^\dagger=0,
\end{equation}
where $\hat{H}(N_j)=N_j\hat{C_j}$ is defined in terms of the
difference operators $\hat{C_j}$. Since each term in the previous expression is
multiplied by $N(x_j)$, which can be any general function, the only possibility
is that each element of the previous series vanishes independently. This leads
to a set of difference equations, one per each $v_j$. Up to an irrelevant 
non-vanishing conformal factor $ (\ell_{\rm Pl}^3k_j)$, each
difference equation reads
\begin{align}\label{eq:difference-eq}\nonumber
&-f_+(\mu_j-4\rho)\phi_j(\mu_j-4)-f_-(\mu_j+4\rho)\phi_j(\mu_j+4)\\
&+f_0(k_j,k_{j-1},\mu_j,M) \phi_j(\mu_j)=0.
\end{align}
where the
function $\phi(\vec k,\vec \mu,M)$  admits a natural decomposition in factors of the form 
\begin{equation}
\phi(\vec k,\vec \mu,M)=\prod_{j=1}^V\phi_j(\mu_j),
\end{equation}
with $\phi_j(\mu_j)=\phi_j(k_j,k_{j-1},\mu_j,M)$.
Then, the Hamiltonian constraint only mixes states with support in lattices of
the labels $\mu_j$, with $j=1,2,\ldots$ $, V,$ of step $4\rho$. In addition,
since the functions $f_\pm(\mu_j)$ vanish in the intervals $[0,\mp2\rho]$,
respectively, different orientations of the labels $\mu_j$ are decoupled. Then,
the solution states belong to the subspaces with support on the semilattices
$\mu_j=\epsilon_j\pm 4\rho {\rm n}_j$, with ${\rm n}_j\in \mathbb{N}$ and
$\epsilon_j\in(0,4\rho]$. In consequence, the constraint only relates states
belonging to separable subspaces of the kinematical one, that we will call
${\cal H}_{\vec{\epsilon}}^B=\bigotimes_{j=1}^V{\cal H}_{\epsilon_j}^B$ in the following.
Analogously to what happens in loop quantum cosmology \cite{mmo,fmo}, it 
suffices to provide the value of $\phi_j(\mu_j=\epsilon_j)$ in order to obtain
the function $\phi_j(\mu_j)$ at any other triad section employing the previous
difference equation.  Moreover, in the limit $\mu_j\to\infty$ the solutions
satisfy the  differential equation 
\begin{align}\label{eq:asympt-diff-op}
&-4\mu_j\partial_{\mu_j}^2\phi-4\partial_{\mu_j}\phi-\frac{\tilde\lambda-1}{\mu_j}\phi +\tilde\omega\mu_j\phi=0
\end{align}
with 
\begin{align}\nonumber
&\tilde\lambda=\frac{3}{4}+(k_j-k_{j-1})^2,\\\label{eq:lam-ome-coef}
&\tilde\omega=\left(1-\frac{2GM}{\ell_{\rm Pl}|k_j|^{1/2}}\right)
\end{align}
which corresponds to either a Bessel or a modified Bessel differential equation
depending if $\tilde\omega$ is negative or positive (as we already indicated in Sec. \ref{sec:prescA} we will refer to these two different situations as the interior or exterior of the
black hole, respectively). The immediate consequence is that the solutions to
the constraint are different depending on whether we are inside or outside the horizon,
and they have to be analyzed separately.

This behavior of the difference equation, together with the numerical studies
carried out on flat and closed Friedmann-Robertson-Walker spacetimes in loop
quantum cosmology (see Ref.\cite{mmo,cfrw}) will allow us to anticipate several
aspects of the solutions to the difference equation~\eqref{eq:difference-eq},
without solving it explicitly. In the next two subsections and in
App.~\ref{app-C} we provide a discussion about this point, but let us summarize
the main results: i) whenever $\ell_{\rm Pl}|k_j|^{1/2}>2GM$ (exterior of the
black hole) we find that the constraint equation can be diagonalized as
$\lambda_n(\epsilon_j)-\Delta k_j^2=0$, where $\{\lambda_n\}$ is a countable set
of positive real numbers that depends on $\epsilon_j\in(0,4\rho]$ associated
with the different discretizations, and for a given $\epsilon_j$, it is expected
that the sequence $\{\lambda_n\}$ will depend as well on $M$ and $k_j$. The
solutions, as functions of $\mu_j$, emerge out of the minimum triad section
$\mu_j=\epsilon_j$, oscillate several times, and decay exponentially in the
limit $\mu_j\to\infty$. This situation can be identified with the one already
found in closed homogeneous and isotropic spacetimes~\cite{cfrw}. ii) On the
other hand, if $\ell_{\rm Pl}|k_j|^{1/2}<2GM$ (interior of the black hole),  the
constraint equation in its diagonal form is $\omega_j-(1-2GM/\ell_{\rm
Pl}|k_j|^{1/2})=0$, where  $\omega_j\in \mathbb{R}^+$. Again, the solutions
emerge out of a minimum triad section, and behave in the limit $\mu_j\to\infty$
as an exact standing wave, i.e., a linear combination of two in and out plane
waves in $\mu_{j}$ of frequency $\omega_j^{1/2}/2$. This situation is analogous
to the one found in Ref.~\cite{mmo} for flat, homogeneous and isotropic
cosmologies.

In both cases, for a given value of either $\lambda_n(\epsilon_j)$ or $\omega_j$, the
corresponding solutions are non-degenerated. Let us see all this in more detail.

\subsubsection{The exterior of the black hole: $\ell_{\rm Pl}|k_j|^{1/2}>2GM$}\label{sec:exterior-B}

Let us recall that, for any choice of $\epsilon_j$, the solutions $\phi_j$ are
completely determined by their initial data $\phi_j(\mu_j=\epsilon_j)$. If we
fix it to be real, together with the fact that the coefficients of the
corresponding difference equation are also real, we conclude that at any other
triad section the corresponding solution $\phi_j(\mu_j)$ will remain a real
function. We will also introduce a bijection on the space of solutions that will
allow us to achieve a suitable separable form of the constraint equation. If we
recall that the zero volume states have been decoupled, the bijection is defined
by the scaling of the solutions
\begin{equation}\label{eq:scaling-out}
\phi^{\rm out}_j(\mu_j)=\hat b(\mu_j)\phi_j(\mu_j),
\end{equation}
with 
\begin{align}\label{eq:bmu-coeff}
\hat b(\mu_j) =
\frac{1}{\rho}(|\hat\mu_j+\rho|^{1/2}-|\hat\mu_j-\rho|^{1/2}),
\end{align}
the square root of the inverse triad operator $\widehat{\left[1/{\mu_j}\right]}$ (up to a factor $\ell_{\rm Pl}^{-1}$), which has an empty kernel.  The new functions $\phi^{\rm out}_j(\mu_j)$ are the coefficients of the solutions to the difference operators 
\begin{equation}
\hat C_j^{\rm
out}=\widehat{\left[\frac1{\mu_j}\right]}^{-1/2}\hat C_j\widehat{\left[\frac1{\mu_j}\right]}^{-1/2}.
\end{equation}
In this situation,
the constraint equation admits a separation of the form
\begin{equation}
\hat C_j^{\rm
out}=\hat{\cal C}_j^{\rm out}-(k_j-k_{j-1})^2,
\end{equation}
where $\hat{\cal C}_j^{\rm out}$ is a difference operator, for each $j$, whose spectral decomposition can be carried out. For consistency, we will study its positive spectrum by solving the eigenvalue problem
\begin{equation}
\hat{\cal C}_j^{\rm out}|\phi^{\rm out}_{\lambda_j}\rangle=\lambda_j|\phi^{\rm out}_{\lambda_j}\rangle.
\end{equation}

In App.~\ref{app-C} we have included a detailed discussion on what we expect
about the  spectrum and the eigenfunctions of these operators. In particular, we
conclude that the eigenvalues belong to a countable set that will depend on the
particular semilattice where the eigenfunction has support, i.e., on 
$\epsilon_j\in(0,4\rho]$ (usually called superselection sector in the loop
quantum cosmology literature), and for a given $\epsilon_j$ the eigenvalues are
expected to depend on $k_j$ and $M$. Therefore, the difference operators $\hat
C_j^{\rm out}$, for a given integer $n$, can be diagonalized as
\begin{equation}\label{eq:scalar-constr-2}
\lambda_n(M,k_j,\epsilon_j)-\Delta k_j^2=0.
\end{equation}

This condition appears to considerably restrict the possible values of $\Delta
k_j$ in the exterior of the black hole. Even one can think seriously in possible
inconsistencies owing to the fact that, for a given choice of $M$,
Eq.~\eqref{eq:scalar-constr-2} cannot be satisfied exactly for all $j$, since
both addends take discrete values without any a priori relation. However, we can
take advantage of the dependence of $\lambda_n(\epsilon_j)$ on the parameter
$\epsilon_j$. The results showed in Ref.~\cite{cfrw} indicate that the
dependence is continuous and in such a way that one can cover the whole positive
real line (up to some interval $[0,\lambda_0]$) with the set
$\{\lambda_n(\epsilon_j)\}$ for each $j$. Therefore, we expect that, for any
given choice of $M$ and $\Delta k_j$, we will be able to find $\epsilon_j$ and
$\lambda_n(\epsilon_j)$ fulfilling Eq.~\eqref{eq:scalar-constr-2}, up to a
region $[0,\lambda_0]$ that must be analyzed carefully.  Since the prescription
we are adopting for the difference operator is different from the one chosen in
Ref.~\cite{cfrw}, we expect that our choice will successfully provide a
satisfactory description also in that region. An interesting question would be
whether this dependence on $\epsilon_j$, apart from continuous, is also
monotonous. If it is discontinuous, there could be values of $\Delta k_j$ that
could not be properly covered. If there is a non-monotonous dependence of
$\lambda_n(\epsilon_j)$ on  $\epsilon_j$ it would be possible to find several
values of $\epsilon_j$ that would be associated to the same eigenvalue
$\lambda_n(\epsilon_j)$, which would allow us to identify new superselection
sectors in the theory (as the usual ones found in loop quantum cosmology) and
consequently additional genuine quantum observables would emerge. This is
a question that will be studied in a future publication.

Let us remark that the corresponding eigenfunctions are normalized to
\begin{equation}
\langle \phi^{\rm out}_{\lambda_n(\epsilon_j)}|\phi^{\rm out}_{\lambda_{n'}(\epsilon_j')}\rangle=\delta_{nn'}\delta_{jj'},
\end{equation}
on the corresponding ${\cal H}_{\epsilon_j}^B$. 

Finally, the role played by $\epsilon_j$ is analogous to the one of the
parameter $\alpha_j$ introduced in App.~\ref{app-B}, where the latter allows one
to cover the whole real line with the spectrum of the momentum operator on a box
$(0,L_j]$.


\subsubsection{The interior of the black hole $\ell_{\rm Pl}|k_j|^{1/2}<2GM$}

In the interior of the black hole, i.e. $\ell_{\rm Pl}|k_j|^{1/2}<2GM$, we can
carry out a similar analysis. Let us introduce this alternative invertible scaling
\begin{equation}
\phi^{\rm in}_j(\mu_j)=\hat \mu_j^{1/2}\phi_j(\mu_j),
\end{equation}
where these new coefficients correspond to the solutions of the difference operators $\hat C_j^{\rm in}=\hat{\mu}_j^{-1/2}\hat C_j\hat{\mu}_j^{-1/2}$. This time, the constraint equations now read
\begin{equation}\label{eq:in-eigen-eq}
\hat{\cal C}_j^{\rm in}+\left(1-\frac{2GM}{\ell_{\rm Pl}|k_j|^{1/2}}\right)=0,
\end{equation}
where $\hat{\cal C}_j^{\rm in}$ is also a difference operator for each $j$ that we will be able to diagonalize after solving the eigenvalue problem 
\begin{equation}
\hat{\cal C}_j^{\rm in}|\phi^{\rm in}_{\omega_j}\rangle=\omega_j|\phi^{\rm out}_{\omega_j}\rangle,
\end{equation}
for $\omega_j\geq 0$. In App.~\ref{app-C} we find that this counterpart of the
spectrum is continuous and non-degenerated. The eigenfunctions are normalized on
${\cal H}_{\epsilon_j}^B$ to 
\begin{equation}
\langle \phi^{\rm in}_{\omega_j}|\phi^{\rm in}_{\omega'_j}\rangle=\delta\left(\sqrt{\omega_j}-\sqrt{\omega_j'}\right).
\end{equation}
In this case, the constraint equations $\hat C_j^{\rm in}$  acquire the diagonal form
\begin{equation}\label{eq:scalar-constr-3}
\omega_j+\left(1-\frac{2GM}{\ell_{\rm Pl}|k_j|^{1/2}}\right)=0.
\end{equation}
which can be satisfied for any $M$ and $k_j$ (compatible with the interior of
the black hole), owing to the continuity of the eigenvalues $\omega_j$.

\subsubsection{Physical Hilbert space}

The solutions to the constraint can be computed applying group averaging. In particular, since the vanishing eigenvalue of the constraint belongs to the continuous spectrum (in the constraint equation $M$ is continuous), the group averaging is
\begin{align}\label{gen-phys-stateB}
&(\Psi^{C}_g|=\int_{-\infty}^{\infty}d\lambda_1\cdots \int_{-\infty}^{\infty}d\lambda_V \; e^{i \sum_j \lambda_j \hat C_j^\dagger}\int_0^\infty dM\sum_{\vec k}\sum_{\vec \mu}\langle g, \vec{k},\vec{\mu},M|\psi(M)\chi(\vec k)\phi(\vec k;\vec{\mu};M), 
\end{align}
assuming the selfadjointness of the constraints $\hat C_j$. Besides, if the corresponding solutions $\{\phi^{\rm out}_{\vec{\lambda}_n}\}$ and $\{\phi^{\rm in}_{\vec{\omega}}\}$ provide a basis (possibly generalized) of the kinematical Hilbert space, the previous group averaging yields
\begin{widetext}
\begin{align}
&(\Psi^{C}_g|=\int_0^{\infty} dM\sum_{\vec{k}<(2GM/\ell_{\rm Pl})^2}\psi(M)\chi(\vec{k})\langle M|\langle \vec{k} |\langle \phi^{\rm in}_{\vec{\omega}(\vec{k},M)} |+\int_0^{\infty} dM\sum_{\vec{k}>(2GM/\ell_{\rm Pl})^2}\psi(M)\chi(\vec{k})\langle M|\langle \vec{k} |\langle \phi^{\rm out}_{\vec{\lambda}_n(\vec{k},M)} |,
\end{align}
\end{widetext}
This formal expression of the solutions provides the Kucha\v{r} mass function
together with  the functions that solve the scalar constraint. In order to
identify a suitable inner product, let us introduce the representation
associated with the canonically conjugate variable to $M$, which we will call
$\tau$, and identify it  with a relational time. We then pick out any arbitrary vertex $v_j$, for instance on the exterior and we solve $M$ in favor of the corresponding eigenvalue $\omega_j$. The solutions are then

\begin{widetext}
\begin{align}
&\Psi^{C}_g(\vec{k},\vec{\mu};\tau)=\frac{2G}{\ell_{\rm Pl}\sqrt{k_j}}\int_0^{\infty} d\omega_j\psi(\omega_j)\chi(\vec{k}) \phi^{\rm in}_{\vec{\omega}(\omega_j)}(\vec{\mu})e^{iM(\omega_j)\tau}+\sum_{\vec{\lambda}_n(\omega_j)}\psi(\vec{\lambda}_n(\omega_j))\chi(\vec{k}) \phi^{\rm out}_{\vec{\lambda}_n(\omega_j)}(\vec{\mu})e^{iM(\vec{\lambda}_n(\omega_j))\tau},
\end{align}
\end{widetext}
recalling that for the exterior, the parameters $\epsilon_j$ cannot be freely chosen, while in the interior they are unconstrained. Besides, the remaining eigenvalues $\omega_j'$ and $\lambda_n(\epsilon_j')$, with $j'\neq j$, are all determined for a given choice of $M$, and consequently of $\omega_j$. Let us also remark that a similar construction can be provided by selecting any vertex on the interior. The solutions have finite norm
\begin{align}
\|\Psi^{C}_g(\tau_0)\|^2=\sum_{\vec{k}}\sum_{\vec{\mu}}|\Psi^{C}_g(\vec{k},\vec{\mu};\tau_0)|^2<\infty.
\end{align}
Therefore, the space of solutions of the scalar constraint is equipped
with a suitable time-independent inner product
\begin{align}
\langle g,\vec{k},\vec{\mu}
|g',\vec{k}',\vec{\mu}'
\rangle=\delta_{\vec{k},\vec{k}'}\delta_{\vec{\mu},\vec{\mu}'}\delta_{g,g'}\;,
\end{align}
which coincides with the one of the spin networks with each $\mu_j$ restricted
to a suitable semilattice $\epsilon_j$. This solution space can be completed
with the previous inner product, which turns out to be
\begin{equation}
{\cal H}^B_{\rm C}=\bigotimes_{j=1}^V\ell^2_j\otimes {\cal H}_{\epsilon_j}^B,
\end{equation} 
the kinematical Hilbert space of the spin networks sector but
restricted to the separable subspaces labeled by~$\vec{\epsilon}$. Finally,
after group averaging these states with the diffeomorphism constraint (see
Sec.~\ref{sec:diffeo-const}), we recover the physical Hilbert space, whose
elements are a superposition of the former with arranged vertices in all
possible positions.

Regarding the observables of the model, it is worth commenting that we can
identify the very same ones found before, that is, the constant of the motion
associated with the mass $\hat M=-i\partial_\tau$ on the boundary and the new
observables $\hat V$ and $\hat O(z)$ on the bulk defined in
Eq.~\eqref{eq:new-obser}.

\section{Diffeomorphism invariant states}\label{sec:diffeo-const}

The physical sector of the system is codified in those states that are invariant
under the symmetries of the model: diffeomorphisms in the radial direction and
time-reparametrizations. These transformations are classically generated by the 
constraints \eqref{eq:abel-class-contr} and \eqref{diff-constr}. In this section we
will deal with the diffeomorphism constraint (for the scalar one see Sec.~\ref{sec:scalar-const}).

The usual strategy followed in loop quantum gravity is to apply the so-called
group averaging technique, which picks out the diffeomorphism invariant states as
well as it induce a natural inner product on the corresponding complex vector
space. One starts with a particular graph $g$ and any element $\Psi_g\in \rm
Cyl$, where ${\rm Cyl}$ is the space of cylindrical functions for all graphs
$g$. One then considers all the diffeomorphisms, and averages each element
$\Psi_g$ with respect to them. The result is all the infinite linear
combinations of diffeomorphism invariant functionals belonging to the algebraic
dual of $\rm Cyl$, i.e. $\rm Cyl^*_{\rm Diff}$. This averaging is made with a suitable rigging map
\begin{align}
\eta: {\rm Cyl}\to {\rm Cyl}^*_{\rm Diff},
\end{align}
that allows one to identify a natural inner product
\begin{align}\label{eq:ga-inner-prod}
\langle \eta(\Psi)|\eta(\Phi)\rangle=\langle \eta(\Psi)|\Phi\rangle,
\end{align}
which is independent of $\Psi$ and $\Phi$. The completion of $\rm Cyl^*_{\rm
Diff}$ with respect to the inner product \eqref{eq:ga-inner-prod} allows the construction of
the Hilbert space of diffeomorphism invariant states of the model, i.e. ${\cal
H}_{\rm Diff}$. This Hilbert space admits a natural decomposition 
\begin{equation}
{\cal H}_{\rm Diff}=\oplus_{[g]}{\cal H}_{[g],\rm Diff},
\end{equation}
where $[g]$ runs over the  diffeomorphism classes of graphs.

In this symmetry reduced model, the diffeomorphism constraint averages states on
the radial direction. Therefore, the resulting space of diffeomorphism invariant
states is given by linear combinations of spin networks with vertices in all
possible positions along the radial line. The corresponding ${\cal H}_{\rm
Diff}$ is endowed with a basis of states that is characterized by the
diffeomorphism class of graphs $[g]$, and each state in a given class by
colorings of edges and vertices.  These states are commonly called
(symmetry-reduced) spin-knot states (or s-knot states). Besides, owing to the
graph symmetry group, the order of the position of the vertices of the
diffeomorphism invariant states must be preserved, i.e., they can be regarded as
an ensemble of chunks of volumes arranged in the radial coordinate. Therefore,
the states which solve the scalar constraint, for both prescriptions of periodic
and quasiperiodic functions, have essentially the same Hilbert space structure
that the one of spin network states. On this space the group averaging can be
carried out in a standard way, yielding the mentioned diffeomorphism invariant
states. Then, since the coloring of the edges $k_j$ is preserved in both
quantization prescriptions, we can identify the observables $\hat V$, the number
of vertex of the graph, and $\hat O(z)$, defined in Eq.~\eqref{eq:new-obser},
with no classical analog.

\section{Semiclassical states}\label{sec:semiclass}

An interesting question is whether we can recover a semiclassical description
out of this quantum theory. The standard procedure consists in looking for those
states where the expectation values of physical observables are peaked on
classical trajectories. For example, a good candidate would be 
\begin{equation}
\psi(M)=\frac{1}{(2\pi\Delta M^2)^{1/4}}e^{-(M-M_0)^2/4(\Delta M_0)^2}, \quad \chi(k_{j'})=\delta_{j'j},\quad k_{j'}>k_{j}\; {\rm if} \; j'>j.
\end{equation}
such that $M_0\gg m_{\rm Pl}$, with $m_{\rm Pl}=\hbar/G$ being the
Planck mass, and $\Delta M_0/M_0\ll1$, where $\Delta M_0$ is the
uncertainly on the mass.  These states may be associated with a
semiclassical description since they provide geometries peaked at a
given mass and around a geometry of a given spin network of coloring
$\vec k$. The consideration of sequences of growing quantum numbers
${k_j}$ is required since they make the radial variable monotonically
growing and avoid double coverings.  In this situation, the areas
of spheres of symmetry are quantized such that the difference of the
areas of two spheres on two arbitrary vertices $v_j$ and $v_{j'}$
would be an integer times a fundamental quanta of area of the order
of the square of the Planck length. Besides, it would be
interesting to consider more general states $\chi(k_j)$. In
this case, the fundamental (but state-dependent) discretization of
the geometry might not correspond to spheres of well-defined,
quantized area, but determined by the specific expectation values
$\langle \vec k\rangle_\chi$.  Therefore, the intrinsic
discretization of the spacetime can be modeled, keeping the
semiclassical character of the effective spacetime. In addition, the
study alternate semiclassical conditions like, e.g., geometries with
a low dispersion on geometrical objects such as the volume
operator
\begin{equation}
\frac{\Delta_{\Psi}\hat {\cal V}}{\langle\hat {\cal V}\rangle_{\Psi}}\ll 1.
\end{equation}
or even the metric components which would depend as well on the connection variables.
These requirements could induce additional restrictions on the physical states.
This will be a matter of research for a future publication.

We would like to remark that on the basis of states labeled by $M$ and the
eigenvalues of $\hat O(z)$  (not necessarily related with semiclassical states)
there exist two regions where the corresponding states show (well defined but)
different behaviors. One of them corresponding to (the square root of the)
eigenvalues of $\hat O(z)$ bigger than $2GM$ and the other one in the opposite
situation. In the cases in which one approximates semiclassical geometries, for
instance when a suitable superposition of masses and $k_j$'s (or certain special
choice for the labels of the states) is considered, these regions would
correspond to the usual notion of exterior and interior of the black hole,
respectively. In these situations, one can study geodesics in the effective
metric in the exterior (defining discrete approximations if necessary) and one
would see that null geodesics end on the asymptotic boundary. This method would
provide an additional notion of what is the exterior region of the
(semiclassical) black hole.

\section{Conclusions}\label{sec:conclus}

In this manuscript we have analyzed the quantization of spherically
symmetric spacetimes adopting loop quantization techniques. In
particular, we consider a canonical description in terms of the real
Ashtekar-Barbero connection and its conjugate variable. After several
canonical transformations and an innocuous gauge fixing, the resulting
model is characterized by two local first class constraints codifying
the invariance under diffeomorphisms in the radial direction and time
reparametrizations. The constraint algebra shows structure functions
that obstruct a subsequent quantization. Fortunately, we can avoid
this obstacle by means of a redefinition of the constraint algebra
after modifying the lapse and the shift functions conveniently, such
that the new constraint algebra is a true Lie algebra. We then apply
the Dirac approach combined with a quantization of the geometry \`a la
loop. Let us recall that the basic bricks of the quantum theory are
holonomies of the Ashtekar-Barbero connection along
piecewise-continuous edges and fluxes of densitized triads through
surfaces. We study two quantization prescriptions where the
corresponding kinematical Hilbert spaces consist in the standard
Kucha\v{r} mass states tensor product with the Hilbert space of spin
networks formed by edges along the radial line joined by vertices
(transverse direction), but for each prescription the point holonomies
are represented as periodic or quasiperiodic functions of the
connection, respectively. In order to solve the dynamics, we start
looking for the solutions to the scalar constraint on each
prescription.  In the case of periodic functions, we study the
quantization already proposed in Ref.~\cite{gp-lett}, where the
solutions can be explicitly obtained. Besides, they belong to a
subspace of the kinematical Hilbert space, and no additional
considerations must be taken into account in order to endow them with
Hilbert space structure. This is no longer the case when we adopt the
alternate prescription for quasiperiodic functions. There, we
represent the scalar constraint on the kinematical Hilbert space
adopting a convenient factor ordering. Its solutions have support on
semilattices of constant step of the triad associated to the
transverse direction. They can be computed out of their initial data
in the minimum volume section. Besides, applying group averaging, the
resulting Hilbert space differs from the kinematical one. In both
cases, we can complete the quantization by implementing the
diffeomorphism constraint after applying group averaging to their
respective solution spaces, achieving different results for each
prescription. However, we were able to identify the very same
observables in the two prescriptions: the traditional Dirac observable
of the model associated to the mass of the black hole, together with a
new one emerging out of both the implementation of the diffeomorphism
constraint and the special properties of the scalar one. The latter
preserves the number of vertices of the states and the former respects
their order. These two facts are the responsible of the emergence of
this new observable, with no classical analog. Therefore, both
descriptions are equivalent at the level of Dirac
observables. Regarding the singularity resolution, we have argued
  in two different ways how it can be avoided in the quantum
  theory. One of them consists in looking for observables, like the
  metric components, suitably defined as evolving
  constants~\cite{cgp,gp-lett}, and require selfadjointness of
  them. In particular, we found strong arguments that such a
  requirement forces the absence of singular
geometries in the quantum theory. The other procedure to reveal
  the resolution of singularities, based on ideas of
  Refs.~\cite{biancI,mmo}, consists in the selection of a suitable
  representation for the scalar constraint such that it has an
  invariant domain free of possible problematic states, and
  consequently restrict the study to this subspace. In this case the
  physical states are linear superpositions of spin networks where the
  triad never vanishes. It in this sense that we claim that the
  classical singularity is not present in the quantum theory.

The obtained results open new possibilities to address fundamental
problems in black hole physics. For instance, the discrete geometry
associated to the radial direction as well as the singularity
resolution could have enlightening consequences in the evaporation
process of a black hole and the information loss paradox
\cite{inf-parax}. Regarding the semiclassical kinematics
\cite{bh-qft-gp}, an infalling observer will cross into the interior
of the black hole in a finite time with respect to an observer at the
spatial infinite.  Since the radial coordinate takes discrete values
and their separation is limited to a minimum value due to the
quantization of the area in loop quantum gravity, the blueshift factor
for an infalling observer viewed from infinity never
diverges. Moreover, the infalling observer will reach the region where
the singularity was expected to be. Nevertheless, the singularity
resolution would allow to continue the geodesics at another spacetime
region. In addition, when coupling a test scalar field with a
semiclassical black hole, and assuming that a quantum field theory on
curved spacetimes will be able to capture to certain extent most of
the relevant physical phenomena of the model, the discretization of
the geometry modifies the predictions with respect to the standard
continuous description. This may help solving several problems of the
latter like the problematic trans-Planckian modes close to the
horizon, which could affect the black hole information paradox
\cite{inf-parax}, and the subsequent approaches like the membrane
paradigm~\cite{membran}, black hole
complementarity~\cite{complemt1,complemt2} or the firewall
phenomenon~\cite{firewall1,firewall2}. All these aspects will require
detailed studies that go beyond the scope of this paper.  Besides, we
also expect in a first approximation that following a similar
treatment like the one provided in Ref.~\cite{aan} for cosmological
perturbations, a test quantum field theory on this quantum spacetime
will in fact experience some and not all the quantum geometry degrees
of freedom, as well as the possible generalizations when the
backreaction of a perturbed model is incorporated, as was done in
Ref.~\cite{fmo} for cosmological settings.

Another interesting extension of all the previous studies concerns the
gravitational collapse \cite{bgms} within loop quantum gravity. In particular,
understanding the complete quantum dynamics of the spherically collapse of a
scalar field in loop quantum gravity would provide the missing ingredients that
would allow us to verify the true nature of black hole evaporation and the black
hole information paradox \cite{ab}. 

Finally, we would like to remark that we have carried out a
quantization of a symmetry reduced model. In this sense, as in any
other similar setting, one has to be careful about interpreting its
physical predictions, since their validity must be trusted only after
confrontation with the full theory.

\section*{Acknowledgements}

We wish to thank Abhay Ashtekar, Luis J. Garay, 
Guillermo A. Mena Marug\'an and Saeed Rastgoo for
comments. This work was supported in part by Grant
No. NSF-PHY-1305000, funds of the Hearne Institute for Theoretical
Physics, CCT-LSU, and Pedeciba.

\appendix

\section{Metric variables, falloff and boundary terms}\label{app-A}

Let us comment on the fact that the phase space variables introduced in Sec.
\ref{sec:class_sys} are related to the metric ones introduced by
Kucha\v{r}~\cite{kuchar} $ds^2=\Lambda^2 dx^2+R^2 d\Omega^2$ by means of the
transformation
\begin{subequations}\label{eq:metric-to-triad}
\begin{align}
&\Lambda=\frac{E^\varphi}{\sqrt{|E^x|}}, &P_\Lambda&= -\sqrt{|E^x|}K_\varphi,\\
&R=\sqrt{|E^x|}, &P_R&=-2\sqrt{|E^x|} K_x -\frac{E^\varphi K_\varphi}{\sqrt{|E^x|}},
\end{align}
\end{subequations}
where $P_\Lambda, P_R$ are the momenta canonically conjugate to $\Lambda$ and
$R$, respectively.

Let us also recall that the falloff conditions of the metric variables were
studied in Ref.~\cite{kuchar}. The maximally extended Schwarzschild spacetime
has two infinities $r\to\pm\infty$, with $r$ the radial coordinate in Kruskal 
variables. Considering also the Kruskal time $t$, we then have 
\begin{align}\nonumber
&\Lambda(t,r)=1+\frac{2GM_{\pm}(t)}{|r|}+{\cal O}(|r|^{-(1+\varepsilon)}),\\\nonumber
&R(t,r)=|r|+{\cal O}(|r|^{-\varepsilon}),\\\nonumber
&P_\Lambda(t,r)={\cal O}(|r|^{-\varepsilon}),\\
&P_R(t,r)={\cal O}(|r|^{-(1+\varepsilon)}),
\end{align}
together with lapse and shift functions
\begin{align}\nonumber
&N(t,r)=N_\pm(t)+{\cal O}(|r|^{-\varepsilon}),\\
&N_x(t,r)={\cal O}(|r|^{-\varepsilon}).
\end{align}

Finally, one might realize that the previous falloff conditions together with
the variations of the action associated to the Hamiltonian defined in
Eq.~\eqref{eq:total-ham} at infinity would give the inconsistent result
$N_\pm(t)=0$. In Ref.~\cite{kuchar} was suggested to include  a boundary
contribution in the action
\begin{equation}
N_+(t)M_+(t)+N_-(t)M_-(t),
\end{equation}
where the functions $N_\pm(t)=\pm\dot\tau_\pm(t)$ might be treated as prescribed
functions,  with $\tau_\pm(t)$ the proper time associated to observers moving
along worldlines of constant radii at  both infinities, respectively. Finally,
on the solution space one can check that  $M_\pm(t)=M$, with $M$ a constant of
the motion corresponding to the mass of the black hole. From this point of
view, this contribution can be interpreted as a global degree  of freedom, since
its contribution to the action is of the form
\begin{equation}
S_{\infty}=\int dt \,\dot{\tau} M,
\end{equation}
whose physical interpretation is an additional term to the symplectic structure.

\section{Selfadjointness of the scalar constraint}\label{app-B}

Let us comment on the fact that, within the quantization prescription compatible with the kinematical Hilbert space ${\cal H}^A_{\rm kin}$, we can require selfadjointness to the scalar constraint whenever we restrict the study to the  exterior of the black hole, i.e., $m_j^2>0$. After the change of variable 
\begin{equation}\label{eq:variab-chang}
dx_j=\frac{m_j dy_j}{4\sqrt{1+m_j^2\sin^2 y_j}},
\end{equation}
which essentially gives
\begin{equation}
x_j(y_j,m_j)=\frac{1}{4}m_jF(y_j,im_j),
\end{equation}
with
\begin{equation}
F(\phi,k)=\int_0^\phi\frac{1}{\sqrt{1-k^2\sin^2t}}dt,
\end{equation}
the Jacobi elliptic integral of the first kind, and the new coordinate $x_j\in\big[0,x_j(2\pi,m_j)\big)$,
the equations in \eqref{eq:diff-eq-prescA} can be written as
\begin{align}
\partial_{x_j}\phi_j=i\omega\phi_j,
\end{align}
which is basically the eigenvalue problem of the momentum operator of a free
particle in box of length $L(m_j)=x_j(2\pi,m_j)$. It is well known that this operator has an infinite number of self-adjoint extensions. In fact, for each $j$, such an extension is characterized by a parameter $\alpha_j\in[0,2\pi)$ whenever one restrict the study to the monoparametric family of dense domains 
\begin{align}
&D_{\alpha_j}(\partial_{x_j})=\Big\{\psi\in {\cal H}: \phi_j\big(L(m_j)\big)=e^{i\alpha_j}\phi_j(0)\quad{\rm and}\quad\langle \phi_j|\partial_{x_j}|\phi_j\rangle<\infty\Big\}.
\end{align}
The spectrum of this operator on each dense domain $\alpha_j$ is equal to
$\omega_n(\alpha_j,m_j)=(2\pi n-\alpha_j)/L(m_j)$, with eigenfunctions
\begin{equation}
\phi_{\alpha_j}(x_j,m_j)=\sqrt{\frac{1}{L(m_j)}}\exp\bigg\{i\frac{(2\pi n-\alpha_j)x_j}{L(m_j)}\bigg\}.
\end{equation}

The constraint equation now reads 
\begin{equation}\label{eq:scalar-diag-gp}
\omega_n(\alpha_j,m_j)-(k_j-k_{j-1})=0.
\end{equation}

We may notice that the parameters $m_j$, $k_j$ and $n$ do not vary continuously.
In consequence, in order to the equation \eqref{eq:scalar-diag-gp} be
consistent, one is forced to select a different self-adjoint extension
$\alpha_j$ such that \eqref{eq:scalar-diag-gp} be satisfied $\forall j$. We then
cannot restrict the study to any arbitrary domain $\alpha_j$ on each $j$. As was
noticed in the alternative quantization prescription, the role of the parameter
$\alpha_j$ can be interpreted with a family of parameters that label the
possible discretizations of the triad $\hat E^\varphi(v_j)$.

\section{Spectral decomposition of difference operators}\label{app-C}

In this appendix we include additional details about the spectral properties of 
the difference operators studied in Sec.~\ref{sec:prescB}. 

\subsection{Difference operator: the exterior}

Let us focus on the difference operator $\hat{\cal C}_j^{\rm out}$ introduced in Sec.~\ref{sec:exterior-B},
which has an action of the form
\begin{align}
&  \hat{\cal C}_j^{\rm out} |\mu_j\rangle=
f^{\rm out}_0(\mu_j,k_j,M) |\mu_j\rangle
-f^{\rm out}_+(\mu_j)|\mu_j+4\rho_j\rangle-f^{\rm out}_-(\mu_j)|\mu_j-4\rho_j\rangle,
\end{align}
with 
\begin{align}
&f^{\rm out}_\pm(\mu_j)=\frac{1}{16\rho^2b(\mu_j)b(\mu_j\pm 4\rho)}|\mu_j|^{1/4}|\mu_j\pm 2\rho|^{1/2}|\mu_j\pm 4\rho|^{1/4}s_{\pm}(\mu_j)s_{\pm}(\mu_j\pm 2\rho),\\\nonumber
&f^{\rm out}_0(\mu_j,k_j,k_{j-1},M)=\frac{\mu_j}{b(\mu_j)^2}\left(1-\frac{2GM}{\ell_{\rm Pl}|k_j|^{1/2}}\right)+\frac{1}{16\rho^2b(\mu_j)^2}\left[({|\mu_j||\mu_j+ 2\rho|)^{1/2}}s_+(\mu_j)s_-(\mu_j+2\rho)\right.\\
&\left.+({|\mu_j||\mu_j- 2\rho|)^{1/2}}s_-(\mu_j)s_+(\mu_j-2\rho)\right],
\end{align}
where $b(\mu_j)$ where defined in Eq.~\eqref{eq:bmu-coeff}. We are interested in  its positive spectrum. Therefore, we will study the solutions to the eigenvalue problem
\begin{equation}
\hat{\cal C}_j^{\rm out}|\phi^{\rm out}_{\lambda_j}\rangle=\lambda_j|\phi^{\rm out}_{\lambda_j}\rangle,
\end{equation}
for $\lambda_j\geq0$. In order to deal with this question, let us start studying the limit $\mu_j\to\infty$. There, the corresponding  difference equations become the differential ones
\begin{equation}\label{eq:mod-bessel-diff-op}
-4\mu_j^2\partial_{\mu_j}^2\phi-8\mu_j\partial_{\mu_j}\phi +\tilde\omega\mu_j^2\phi=\gamma_j^2\phi,
\end{equation}
with $\gamma_j^2 = \lambda_j+1$ and $\tilde\omega$ given in Eq.~\eqref{eq:lam-ome-coef}. These equations correspond to modified Bessel
equations where their solutions are combinations of modified Bessel functions,
i.e.,
\begin{equation}
\phi=A x_j^{-1/2}{\cal K}_{i\gamma_j}\left(x_j\right)+B x_j^{-1/2}{\cal I}_{i\gamma_j}\left(x_j\right),
\end{equation}
with $x_j=\mu_j\sqrt{\tilde\omega}/2$. In the limit $\mu_j\to\infty$, ${\cal I}$ grows exponentially and ${\cal K}$ decays exponentially. Therefore, the latter is the only contribution to the spectral decomposition. In consequence, this
counterpart of the spectrum of the difference operator \eqref{eq:mod-bessel-diff-op} is non-degenerate. Moreover, the functions
${\cal K}_{i\gamma_j}(x)$ are normalized to
\begin{equation}
\langle {\cal K}_{i\gamma_j}|{\cal K}_{i\gamma_j '}\rangle=\delta(\gamma_j-\gamma_j'),
\end{equation}
in $L^2(\mathbb{R},x_j^{-1}dx_j)$, since the normalization in this case is ruled by the behavior of
${\cal K}_{i\gamma}(x)$ in the limit $x\to 0$, which corresponds to
\begin{equation}
\lim_{x\to 0}{\cal K}_{i\gamma_j}(x)\to A\cos\left(\gamma_j\ln|x|\right).
\end{equation}
For additional details, see also Ref.~\cite{kiefer}

Now we appeal to the results found in Ref.~\cite{cfrw}, where the homogeneous
constraint equation is analogous to ours at each vertex $v_j$. There it was found
that the eigenfunctions of such a difference operator have a similar asymptotic
behavior for $\mu_j\to\infty$. However, the spectrum of the corresponding
difference operator turns out to be discrete (instead of continuous like the
corresponding differential operator) owing to the behavior of its eigenfunctions
at $\mu_j\to 0$. Therefore, we expect that $\lambda_n$ belong to a countable set
(for each vertex $v_j$), which must be determined numerically. We expect that
the possible positive values of $\lambda_{n}$ will depend on $\epsilon_j\in(0,4\rho]$, and for a given $\epsilon_j$, they will also depend on $\tilde{\omega}$,
i.e., on $k_j$ and $M$ by means of Eq.~\eqref{eq:lam-ome-coef}.

Therefore, the corresponding eigenfunctions will be normalized on ${\cal H}_{\epsilon_j}^B$ to
\begin{equation}
\langle \phi^{\rm out}_{\lambda_n(\epsilon_j)}|\phi^{\rm out}_{\lambda_{n'}(\epsilon_j)}\rangle=\delta_{nn'}.
\end{equation}

\subsection{Difference operator: the interior}

In addition, we will consider the difference operator $\hat{\cal C}_j^{\rm in}$ that was also introduced in Sec.~\ref{sec:exterior-B},
whose action on a state $|\mu_j\rangle$ is of the form
\begin{align}\nonumber
&  \hat{\cal C}_j^{\rm in} |\mu_j\rangle=
f^{\rm in}_0(\mu_j,k_j,M) |\mu_j\rangle
-f^{\rm in}_+(\mu_j)|\mu_j+4\rho_j\rangle-f^{\rm in}_-(\mu_j)|\mu_j-4\rho_j\rangle,
\end{align}
with 
\begin{align}
&f^{\rm in}_\pm(\mu_j)=\frac{1}{16\rho^2}|\mu_j|^{-1/4}|\mu_j\pm 2\rho|^{1/2}|\mu_j\pm 4\rho|^{-1/4}s_{\pm}(\mu_j)s_{\pm}(\mu_j\pm 2\rho),\\\nonumber
&f^{\rm in}_0(\mu_j,k_j,k_{j-1},M)=\frac{1}{16\rho^2|\mu_j|}\left[({|\mu_j||\mu_j+ 2\rho|)^{1/2}}s_+(\mu_j)s_-(\mu_j+2\rho)\right.\\
&\left.+({|\mu_j||\mu_j- 2\rho|)^{1/2}}s_-(\mu_j)s_+(\mu_j-2\rho)\right]-\frac{{\rm sgn}(\mu_j)}{|\mu_j|\rho^2}(k_j-k_{j-1})^2(|\mu_j+\rho|^{1/2}-|\mu_j-\rho|^{1/2})^2.
\end{align}
We will study then the solutions to 
\begin{equation}
\hat{\cal C}_j^{\rm in}|\phi^{\rm in}_{\omega_j}\rangle=\omega_j|\phi^{\rm out}_{\omega_j}\rangle,
\end{equation}
with $\omega_j\in\mathbb{R}^+$, i.e., we will consider only the positive
counterpart of its spectrum, since it is the only consistent contribution to
Eq.~\eqref{eq:in-eigen-eq}. Let us recall that the coefficients $\phi_j(\mu_j)$
of these difference equations are determined by their initial data
$\phi_j(\mu_j=\epsilon_j)$ and are real if
$\phi_j(\mu_j=\epsilon_j)\in\mathbb{R}$, since the previous functions $f^{\rm
in}_0$ and $f^{\rm in}_\pm$ are also real. We conclude that the eigenfunctions
must be real and the spectrum non-degenerated. Besides, in the limit
$\mu_j\to\infty$, the solutions satisfy the Bessel equations
\begin{equation}\label{eq:bessel-diff-op}
-4\partial_{\mu_j}^2\phi-\frac{\gamma^2}{\mu_j^2}\phi -\omega_j\phi=0,
\end{equation}
whose solutions can be split in linear combinations of Hankel functions of first $H^{(1)}_{i\gamma}(x_j)$ and second kind  $H^{(2)}_{i\gamma}(x_j)$, multiplied by a factor $x_j^{1/2}$, that is
\begin{equation}
\phi=Ax^{1/2}H^{(1)}_{i\gamma}(x_j)+Bx^{1/2}H^{(2)}_{i\gamma}(x_j),
\end{equation}
where
\begin{align}
x_j=\frac{\mu_j\sqrt{\omega_j}}{2},
\end{align}
and $\gamma^2=\tilde\lambda$, with $\tilde{\lambda}$ given in Eq.~\eqref{eq:lam-ome-coef}. The asymptotic limit of these functions is
\begin{equation}
H^{(1)}_{i\gamma}(x)=\sqrt{\frac{2}{\pi x}}e^{i(x-\pi/4+\gamma\pi/2)},\quad H^{(2)}_{i\gamma}(x)=\big(H^{(1)}_{i\gamma}(x)\big)^*.
\end{equation}
This allows us to conclude that, since $\phi^{\rm in}_j(\mu_j)\in\mathbb{R}$, the most
general solution at $\mu_j\to\infty$ must be of the form
\begin{equation}\label{eq:eigen-int-prescB}
\lim_{\mu_j\to\infty}\phi^{\rm in}_{\omega_j}(\mu_j)=A\cos\left[\frac{\sqrt{\omega_j}}{2}\mu_j+\beta\right],
\end{equation}
with $A$ certain amplitude and $\beta$ a phase that can depend on $k_j$,
$k_{j-1}$ and $\epsilon_j$.  We then conclude that the solutions, as functions
of $\mu_j$,  are normalizable (in the generalized sense) to
\begin{equation}
\langle \phi^{\rm in}_{\omega_j}|\phi^{\rm in}_{\omega'_j}\rangle=\delta\left(\sqrt{\omega_j}-\sqrt{\omega_j'}\right),
\end{equation}
on ${\cal H}_{\epsilon_j}^B$.

\end{document}